\documentclass[utf8]{frontiersFPHY} 

\usepackage{url,hyperref,microtype,subcaption}
\usepackage[onehalfspacing]{setspace}

\usepackage{graphics}
\usepackage{graphicx}
\usepackage{bm}
\usepackage{amsmath}
\usepackage{amsfonts}
\usepackage{amssymb}
\usepackage{latexsym}
\usepackage[dvipsnames]{xcolor}
\usepackage{bbold}
\usepackage{braket}

\usepackage{tikz,tikz-3dplot,bm}
\usepackage{hyperref}
\usepackage{physics}
\usepackage{float}

\def\keyFont{\fontsize{8}{11}\helveticabold }
\def\firstAuthorLast{Dawson {et~al.}} 
\def\Authors{Benjamin Dawson\,$^{1,2,*}$, Nicholas Furtak-Wells\,$^{2}$, Thomas Mann\,$^{1}$, Gin Jose\,$^{1}$ and Almut Beige\,$^{2}$}
\def\Address{$^{1}$School of Chemical and Process Engineering, University of Leeds, Leeds LS2 9JT, United Kingdom \\
$^{2}$School of Physics and Astronomy, University of Leeds, Leeds LS2 9JT, United Kingdom}
\def\corrAuthor{Benjamin Dawson}

\def\corrEmail{py13bhd@leeds.ac.uk}

\begin{document}
\onecolumn
\firstpage{1}

\title[Asymmetric mirrors with coherent absorption]{The quantum optics of asymmetric mirrors with coherent light absorption} 

\author[\firstAuthorLast ]{\Authors} 
\address{} 
\correspondance{} 

\extraAuth{}

\maketitle

\begin{abstract}
The local observables of the quantised electromagnetic field near a mirror-coated interface depend strongly on the properties of the media on {\em both} sides. In macroscopic quantum electrodynamics, this fact is taken into account with the help of optical Green's functions which correlate the position of an observer with all other spatial positions and photon frequencies. Here we present an alternative, more intuitive approach and obtain the local field observables with the help of a quantum mirror image detector method [Furtak-Wells {\em et al.}, Phys.~Rev.~A {\bf 97}, 043827 (2018)]. In order to correctly normalise electric field operators, we demand that spontaneous atomic decay rates simplify to their respective free space values far away from the reflecting surface. Our approach is interesting, since mirror-coated interfaces constitute a common basic building block for quantum photonic devices.

\tiny
 \keyFont{ \section{Keywords:} quantum photonics, quantum optics, macroscopic quantum electrodynamics, open quantum systems, spontaneous photon emission}
\end{abstract}

\section{Introduction}

The fluorescence properties of an atomic dipole depend primarily on the so-called {\em local density of states} of the electromagnetic (EM) field, i.e.~on the number of EM mode decay channels available at the same location \cite{Bart,Bart2,Kwarin}. For example, inside a homogeneous dielectric medium with refractive index $n$, the spontaneous decay rate $\Gamma_{\rm med}$ of an atomic dipole equals \cite{glauber,scheel}
\begin{eqnarray} \label{1}
\Gamma_{\text{med}} &=& n \, \Gamma_{\rm air} 
\end{eqnarray}
to a very good approximation, where $\Gamma_{\rm air}$ denotes the corresponding free space decay rate. However, deriving the local density of states of the EM field in more complex scenarios, which involves the calculation of the imaginary parts of the dyadic Green's function \cite{Hecht,Buhmann,Stourm,Bennett2}, can be computationally challenging. Although such calculations can aid the design of photonic devices, they do not provide much physical intuition.

 Taking a different approach, Carniglia and Mandel \cite{carniglia} modelled semi-transparent mirrors by only considering stationary photon modes which contain incoming as well as reflected and transmitted contributions. Their so-called triplet modes  depend on reflection and transmission rates and are a subset of the free space photon modes of the EM field. Unfortunately, this approach can result in the prediction of unphysical interference effects when modelling light approaching a mirror from both sides \cite{zakowicz}. If one wants to avoid such interference problems, adjustments have to be made \cite{creatore,khosravi,furtak,southall}, for example by doubling the usual Hilbert space of the quantised EM field in the presence of a semi-transparent mirror \cite{furtak}. However, this immediately raises the question where the doubling of the Hilbert space comes from. For a detailed discussion of this question see a recent paper by Southall {\em et al.}~\cite{southall} which models two-sided semi-transparent mirrors with the help of locally-acting mirror Hamiltonians and a recent paper by Hodgson {\em et al.}~\cite{hodgson} which quantises the electromagnetic field in position space.

\begin{figure}[t]
	\centering
	\includegraphics[width=8.6cm]{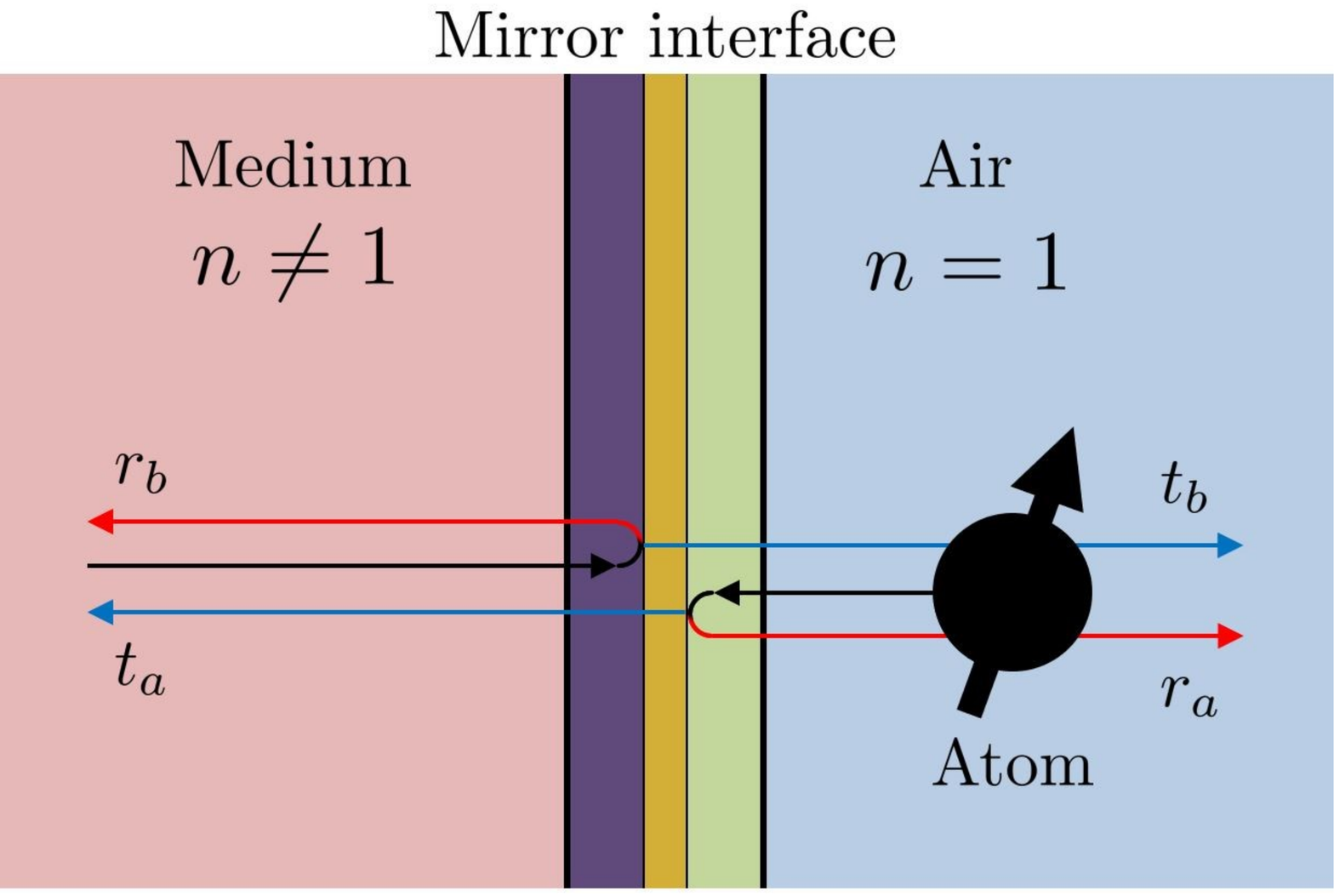}
	\caption{ Schematic view of a mirror-coated dielectric medium with air on its right hand side. The coating which can be characterised by its electric field reflection and transmission rates $r_a$, $r_b$, $t_a$ and $t_b$ may consist of different layers and materials. The possible absorption of light in the interface is explicitly taken into account when we derive the basic observables of the quantised EM field with the help of a quantum mirror image detector method \cite{furtak}. In order to correctly normalise field operators, we introduce a test atom and demand that its spontaneous decay rate simplifies at large distances to the respective free space expression.}
	\label{figpaperlogo}
\end{figure} 

In the following we use the quantum mirror image detector method which has recently been introduced by Furtak-Wells {\em et al.}~\cite{furtak} to obtain the basic observables of the quantised EM field in the presence of  a mirror-coated dielectric interface. This method maps light scattering in the presence of a two-sided semitransparent mirror onto two analogous free space scenarios. More concretely, in our model, we choose an initial time $t=0$ and use one Hilbert space (labelled $a$) to describe the EM field on the right and another one (labelled $b$) to describe the EM field on the left hand side of the mirror interface. For times $t>0$, we assume that their state vectors evolve simply as they would in free space. To identify the electric field amplitude seen by a detector at a certain position ${\bf r}$ and a given time $t$ in the experimental setup in Fig.~\ref{figpaperlogo}, we notice that this amplitude is a superposition of electric field amplitudes seen in the two corresponding free space scenarios. To construct the electric field observable for the above experimental setup, we therefore sum up the signals seen by the original detector and a mirror image detector after placing them in the corresponding free space scenarios. Doubling the Hilbert space of the EM field and distinguishing two different types of photons, namely $a$ and $b$ photons, helps to ensure that wave packets which never meet in real space do not interfere in our model.

The experimental setup which we consider here consists of a dielectric medium with refractive index $n \neq 1$, a mirror coating and air with refractive index $n=1$ next to the coating, as illustrated in Fig.~\ref{figpaperlogo}. The possible absorption of light in the mirror interface, which may consist of different layers and may contain different materials, is explicitly taken into account. However, for simplicity, we only consider coherent light absorption and assume that  incoming wave packets do not lose their coherence properties when passing through the interface. In this case, there is a linear relation between incoming and outgoing electric field amplitudes which allows us to characterise the mirror interface by (real) electric and magnetic field reflection and transmission rates $r_a$, $r_b$, $t_a$ and $t_b$. Moreover, the complex amplitudes of electric field vectors accumulate phase factors $\phi_1$, ..., $\phi_4$ upon reflection and transmission. The indices $a$ and $b$ refer to light approaching the mirror from the left and from the right hand side, respectively. 

In the absence of losses, energy is conserved and Stokes relation implies that the reflection rates for both sides of the mirror interface are the same $(r_a = r_b)$. In addition, the phases $\phi_i$ obey certain conditions \cite{phase1,zeilinger}. However, suppose losses are taken into account and the absorption rate for light approaching the reflecting layer of the mirror interface from the left is much higher than the absorption rate for light approaching from the right. In this case, the reflection rate $r_b$ is much smaller than $r_a$, even for a symmetric reflecting layer, and Stokes relation no longer applies. Instead, for mirror interfaces with coherent light absorption, we have $r_a \neq  r_b$ \cite{Monzon,Jeffers,Pinkse}. In the literature, interfaces with this property are usually referred to as asymmetric mirrors, since they break the forward-backward scattering symmetry of conventional semi-transparent mirrors \cite{schwanecke,plum,zhukovsky,kitur,lezec,kenanakis,filonov}. An alternative way of breaking the symmetry of ideal mirrors, i.e.~without the introduction of absorbing layers, is to use surface roughness. Suppose, the reflecting layer is very smooth and highly-reflecting on one side but diffracts light on the other, then we also effectively have $r_a \neq  r_b$. 

In the following, we construct the observables of the quantised EM field near a mirror-coated interface with coherent light absorption. To correctly normalise these observables, we demand locality and assume that the spontaneous decay rate of a test atom at a relatively large distance $x$ from the reflecting surface equals its free space value. The spontaneous atomic decay rates near highly-reflecting mirrors \cite{morawitz,stehl,milonni,arnoldus, drabe,meschede,amos,matloob,zoller,walther} and near dielectric media with and without losses \cite{carniglia,creatore,khosravi,wylie,snoeks,yeung,rikken, xu, eberlein,wang,falinejad} have already been studied extensively in the literature and theoretical predictions are generally in very good agreement with experimental findings \cite{drexhage,chance,blatt,creatore2}. Like these papers, we ignore interactions of the atomic dipole with the quantum matter of the mirror surface. Instead we assume here that the test atom and the atoms inside the mirror surface are strongly detuned. For simplicity, we also neglect the angle-dependence of reflection rates. 

Despite taking an alternative approach, our results are in good agreement with previous results. In addition, our approach allows us to model scenarios which are not as easily accessible using alternative approaches. For example, the main difference between the setup considered in Ref.~\cite{furtak} and the setup which we consider here is the presence of a dielectric with $n>1$ on the left hand side of the interface. The main difference between Ref.~\cite{creatore} and our calculations is that we allow for arbitrary mirror coatings, including asymmetric mirrors and mirrors with coherent light absorption.   

This paper comprises five sections. In Section \ref{sec2} we quantise the EM field in a homogenous medium with a refractive index $n \neq 1$ by mapping this situation onto an analogous scenario with $n=1$. Section \ref{sec3} covers the quantisation of the EM field in the presence of a mirror-coated interface using the mirror image detector method. In Section \ref{sec4} we determine the missing normalisation factors of electric and magnetic field amplitudes by calculating the spontaneous emission rate of a test atom. Lastly, Section \ref{secconclusion} contains a summary of our findings. 

\section{The quantised EM field inside a dielectric medium} \label{sec2}

The purpose of this section is to obtain the Hamiltonian and the electric and magnetic field observables of the quantised EM field inside a dielectric medium with refractive index $n$. To do so, we relate its properties to the properties of the quantised EM field in an analogous free space scenario.

\subsection{Maxwell's equations} \label{sectionIIA}

Our starting point is classical electrodynamics. In a dielectric medium with permittivity $\varepsilon$ and permeability $\mu$ and in the absence of any charges and currents, Maxwell's equations state that \cite{stratton} 
\begin{eqnarray}  \label{Maxwellmed}
&& \hspace*{-0.5cm} \mathbf{\nabla} \cdot \mathbf{E}_{\text{med}}(\mathbf r,t) = \mathbf{\nabla} \cdot \mathbf{B}_{\text{med}}(\mathbf r,t) = 0 \, , \nonumber \\
&& \hspace*{-0.5cm} \mathbf{\nabla} \times \mathbf{E}_{\text{med}}(\mathbf r,t) = - {\bf \dot B}_{\rm med}({\bf r},t) \, , \nonumber \\
&& \hspace*{-0.5cm} \mathbf{\nabla} \times \mathbf{B}_{\text{med}}(\mathbf r,t) = \varepsilon \mu \, {\bf \dot E}_{\rm med}({\bf r},t) \, .
\end{eqnarray}
Here $\mathbf{E}_{\text{med}}(\mathbf r,t)$ and $\mathbf{B}_{\text{med}}(\mathbf r,t)$ denote electric and magnetic field vectors at positions ${\bf r}$ and times $t$. Moreover, we know that the energy of the EM field inside the dielectric medium equals
\begin{eqnarray} \label{fieldhamiltonianmedium}
H_{\rm med} &=& {1 \over 2} \int_{\mathbb{R}^3} {\rm d}^3{\bf r} \left[ \varepsilon {\bf E}_\text{med}({\bf r},t)^2 + \frac{1}{\mu} {\bf B}_\text{med}({\bf r},t)^2 \right] \, . ~~
\end{eqnarray}
As an example, we now have a closer look at horizontally polarised light which propagates along the $x$-axis. In this case, consistency with Maxwell's equations and with the right hand rule of classical electrodynamics requires that ${\mathbf{E}}_{\rm med}({\mathbf{r}},t) = (0,E_{\rm med}({x},t),0)$ and ${\mathbf{B}}_{\rm med}({\mathbf{r}},t) = (0,0,B_{\rm med}({x},t))$ for wave packets travelling in the positive $x$ direction. Moreover, ${\mathbf{E}}_{\rm med}({\mathbf{r}},t) = (0,{E}_{\rm med}( x,t),0)$ and ${\mathbf{B}}_{\rm med}({\mathbf{r}},t) = (0,0,-B_{\rm med}(x,t))$ for wave packets travelling in the negative $x$ direction. Substituting these vectors into Eq.~(\ref{Maxwellmed}), they reduce to the differential equations
\begin{eqnarray} \label{maxwellsolution+x}
{\partial_x{B}_{\rm med}(x,t)} &=& \pm\varepsilon \mu \, {\partial_t {E}_{\rm med}(x,t)} \, , \nonumber \\  
{\partial_x {E}_{\rm med}(x,t)} &=& \pm {\partial_t {B}_{\rm med}(x,t)} \, ,
\end{eqnarray}  
where the minus and plus signs correspond to different directions of propagation. The solutions of these equations are wave packets which travel at the speed of light $c = 1/  \sqrt{\varepsilon \mu}$. Analogous equations apply for vertically-polarised light travelling along the $x$ axis and for light travelling in other directions. 

A special example of a dielectric medium is air with $\varepsilon = \varepsilon_0$ and $\mu = \mu_0$. In the following, we denote the corresponding field vectors by ${\bf E}_{\text{air}}({\bf r},t)$ and ${\bf B}_{\text{air}}({\bf r},t)$. Using this notation, a closer look at Eq.~(\ref{maxwellsolution+x}) implies the equivalency relations
\begin{eqnarray} \label{equivalence}
{\bf E}_{\text{med}}({\bf r},t) &=& \sqrt{n^3 \varepsilon_0 \over \varepsilon} \, {\bf E}_{\rm air}\left(n {\bf r} ,t \right) \, , \nonumber \\
{\bf B}_{\text{med}}({\bf r},t) &=& \sqrt{n^3 \mu \over \mu_0} \, {\bf B}_{\rm air}\left(n {\bf r}, t \right) 
\end{eqnarray}
with the refractive index, as usual, defined as
\begin{eqnarray} \label{defn2}
n &=& \sqrt{ \varepsilon \mu \over \varepsilon_0 \mu_0} \, . 
\end{eqnarray}
For air, we simply have $n=1$. Eq.~(\ref{equivalence}) guarantees that ${\bf E}_{\text{med}}({\bf r},t)$ and ${\bf B}_{\text{med}}({\bf r},t)$ solve Maxwell's equations in a dielectric medium when ${\bf E}_{\rm air}\left(n {\bf r} ,t \right)$ and ${\bf B}_{\rm air}\left(n {\bf r} ,t \right)$ solve Maxwell's equations in air. 

One difference between electric and magnetic field solutions in a dielectric medium and in air is a re-scaling of field vector amplitudes. Here the factors on the right hand side of Eq.~(\ref{equivalence}) have been chosen such that $H_{\rm med}$ in Eq.~(\ref{fieldhamiltonianmedium}) and the energy $H_{\rm air}$ of the EM field in air, 
\begin{eqnarray} \label{fieldhamiltonianmedium4}
H_{\rm air} &=& {1 \over 2} \int_{\mathbb{R}^3} {\rm d}^3{\bf r} \left[ \varepsilon_0 {\bf E}_\text{air}({\bf r},t)^2 + \frac{1}{\mu_0} {\bf B}_\text{air}({\bf r},t)^2 \right] \, , ~~
\end{eqnarray}
are the same,
\begin{eqnarray} \label{oma1}
H_{\rm med} &=& H_{\rm air} \, .
\end{eqnarray}
Moreover, on the right hand side of Eq.~(\ref{equivalence}) there is a re-scaling of the position vector ${\bf r}$. Inside the medium, light travels a shorter distance in the same amount of time but electric and magnetic field amplitudes still oscillate locally at the same rate \cite{stratton,griffiths}.

\subsection{Field quantisation in air} \label{freely}

Wave-particle duality suggests that the EM field is made up of particles, i.e.~photons \cite{bennett}. In the case of light propagation in three dimensions, we characterise each photon by its polarisation $\lambda$ and its wave vector ${\bf k}$. Moreover, we know from experiments that a photon with wave vector ${\bf k}$ has the energy $\hbar \omega$ with $\omega = c_0 \, \| {\bf k} \|$ and $c_0 =  1 / \sqrt{\varepsilon_0 \mu_0}$. Hence the Hamiltonian of the quantised EM field can be written as 
\begin{eqnarray}\label{quantisedfield3D}
H_{\rm air} &=& \sum _{\lambda = \sf H, \, \sf V} \int_{\mathbb{R}^3} {\rm d} \mathbf k \, \hbar \omega \, a^\dagger_{\mathbf k \lambda} a_{\mathbf k \lambda} \, ,
\end{eqnarray}
where $a_{{\bf k} \lambda}$ with the bosonic commutator relation $[a_{\mathbf k \lambda}, a^\dagger_{\mathbf k' \lambda'}] = \delta_{\lambda, \lambda'} \, \delta(\mathbf k -\mathbf k')$ denotes the annihilation operator of photons in the $({\bf k},\lambda)$ mode. Consistency with classical electrodynamics (c.f.~Eq.~(\ref{fieldhamiltonianmedium4})) requires that this Hamiltonian coincides, up to a constant, with the observable 
\begin{eqnarray} \label{fieldhamiltonianmediumx}
H_{\rm air} &=& {1 \over 2} \int_{\mathbb{R}^3} {\rm d}^3{\bf r} \left[ \varepsilon_0 \, {\bf E}_{\rm air}({\bf r})^2 + \frac{1}{\mu_0} \, {\bf B}_{\rm air}({\bf r})^2 \right] \, , 
\end{eqnarray}
where $\mathbf E_{\rm air}(\mathbf r)$ and $\mathbf B_{\rm air}(\mathbf r)$ denote the electric and magnetic field free space observables. Hence both observables are linear superpositions of photon annihilation and creation operators.  Demanding consistency with Maxwell's equations and taking the above field Hamiltonian into account, they can be shown to equal \cite{bennett} 
\begin{eqnarray}\label{EMfieldfree3d}
\mathbf E_{\rm air}(\mathbf r) &=& {{\rm i} \over 4 \pi} \sum _{\lambda = 1,2} \int_{\mathbb R^3} {\rm d}^3 \mathbf k \, \sqrt{\frac{\hbar \omega}{\pi\varepsilon_0}} \,  \text{e}^{\text{i}\mathbf k \cdot \mathbf r} \, a_{\mathbf k \lambda} \, \hat{\mathbf{e}}_{\mathbf k \lambda} + \text{H.c.} \, ,  \nonumber \\ 
\mathbf B_{\rm air}(\mathbf r) &=& - {{\rm i} \over 4\pi c_0} \sum _{\lambda = 1,2} \int_{\mathbb{R}^3} {\rm d}^3 \mathbf k \, \sqrt{\frac{\hbar \omega}{\pi \varepsilon_0}} \, \text{e}^{\text{i}\mathbf k \cdot \mathbf r} \, a_{\mathbf k \lambda} \, \hat{\mathbf k} \times \hat{\mathbf{e}}_{\mathbf k \lambda} + \text{H.c.} 
\end{eqnarray} 
with $k = \| {\bf k} \|$ and ${\bf \hat k} = {\bf k}/\| {\bf k} \| $. Here $\hat{\mathbf{e}}_{\mathbf k \lambda}$ denotes a polarization vector with $\hat{\mathbf{e}}_{\mathbf k \lambda} \cdot \mathbf {k}=0$ and $\| \hat{\mathbf{e}}_{\mathbf k \lambda} \| =1$. The normalisation factors in Eq.~(\ref{EMfieldfree3d}) have been chosen such that Eqs.~(\ref{quantisedfield3D}) and (\ref{fieldhamiltonianmediumx}) differ only by a constant term with no physical consequences. 

\subsection{Field quantisation in a dielectric medium}

To obtain the electric and magnetic field observables ${\bf E}_\text{med}({\mathbf r})$ and ${\bf B}_\text{med}({\mathbf r})$ inside a dielectric medium, we now map the dynamics of wave packets inside the medium onto analogous free-space dynamics. In other words, we quantise the EM field in the dielectric medium in terms of {\em free space photons}. To do so, we employ the equivalency relations  in Eq.~(\ref{equivalence}) which imply that 
\begin{eqnarray} \label{quantisedfieldmed3D}
{\bf E}_\text{med}({\mathbf r}) &=& \sqrt{\frac{n^3 \varepsilon_0}{\varepsilon}} \, {\bf E}_{\rm air}(n{\mathbf r}) \, , \nonumber \\
{\bf B}_\text{med}({\mathbf r}) &=& \sqrt{\frac{n^3 \mu}{\mu_0}} \, {\bf B}_{\rm air}(n{\mathbf r}) 
\end{eqnarray} 
with ${\bf E}_{\rm air}({\mathbf r})$ and ${\bf B}_{\rm air}({\mathbf r}) $ given in Eq.~(\ref{EMfieldfree3d}). From Eq.~(\ref{fieldhamiltonianmedium4}) we see that the energy observable of the EM field in a dielectric medium equals
\begin{eqnarray} \label{fieldhamiltonianmedium4x}
H_{\rm med} &=& {1 \over 2} \int_{\mathbb{R}^3} {\rm d}^3{\bf r} \left[ \varepsilon {\bf E}_\text{med}({\bf r})^2 + \frac{1}{\mu} {\bf B}_\text{med}({\bf r})^2 \right] \, .
\end{eqnarray}
Using this equation, one can show that the EM field Hamiltonian of the dielectric medium and $H_{\rm air}$ in Eq.~(\ref{quantisedfield3D}) are the same,
\begin{eqnarray} \label{changeinwavevector}
H_{\rm med} &=& H_{\rm air} \, ,
\end{eqnarray} 
as suggested by Eq.~(\ref{oma1}). In our description, a photon of frequency $\omega$ has the energy $\hbar \omega$ in the medium and in free space. The only expectation value that changes when we consider a wave packet of light inside a dielectric medium instead of considering the same quantum state in free space are its electric and magnetic field expectation values. Our ability to describe the dielectric medium with the help of free space observables becomes important in the next section, when we quantise the EM field in the presence of a mirror-coated dielectric medium.

\section{The quantised EM field in the presence of a mirror-coated interface} \label{sec3}

To determine the field Hamiltonian $H_\text{mirr}$ of the quantised EM field in Fig.~\ref{figpaperlogo}, we only consider {\em free space photons travelling in air}. As usual, we characterise each photon by its polarisation $\lambda$ and by its wave vector ${\bf k}$ and assume that its energy equals $\hbar \omega $ with $\omega = c \| {\bf k} \|$. However, as mentioned already in the Introduction, in the presence of the mirror interface, we need to double the Hilbert space of the quantised EM field. In the following, we therefore consider two Hilbert spaces which we label $a$ and $b$ and which describe light on the right and light on the left hand side of the mirror surface, respectively, at a given time $t=0$. Describing both sides separately helps us later on to identify how field excitations contribute to local electric and magnetic field observables \cite{furtak}. Hence $H_{\rm mirr}$ equals
\begin{eqnarray}\label{systemhamiltonian2}
H_\text{mirr} &=&\sum_{s = \pm 1} \sum_{\lambda = 1,2} \int_{\mathbb R ^3} {\rm d}^3 \mathbf k \, \hbar \omega \left[ a_{\mathbf k \lambda}^\dagger a_{\mathbf k \lambda} + b_{\mathbf k \lambda}^\dagger b_{\mathbf k \lambda} \right] , ~~~
\end{eqnarray}
where $a_{\mathbf k \lambda}$ and $b_{\mathbf k \lambda}$ are bosonic annihilation operators with $[ a_{{\mathbf k} \lambda}, b^\dagger_{{\mathbf k}' \lambda'} ] = 0$ and $[b_{\mathbf k \lambda}, b^\dagger_{\mathbf k' \lambda'}] = \delta_{\lambda, \lambda'} \, \delta(\mathbf k -\mathbf k')$. Next we derive the corresponding electric field observable ${\bf E}_\text{mirr}({\bf r})$.

\subsection{Highly-reflecting mirrors} \label{subsecmethodofimagesperfectmirror}

However, for simplicity, we first have a closer look at a highly-reflecting mirror. In this case, an incoming wave packet changes its direction of propagation upon reaching the interface such that its angle of incidence equals its angle of reflection. Suppose the mirror is placed in the $x=0$ plane. In this case, the $y$ and the $z$ component of the electric field vectors of the incoming light accumulate a minus-sign upon reflection to ensure that they remain orthogonal to the direction of propagation. Now suppose a detector measures the electric field amplitude at a position ${\bf r} = (x,y,z)$ in the experimental setup shown in Fig.~\ref{figpaperlogo}. Then the mirror image method of classical electrodynamics \cite{furtak} suggests that the electric field seen by the detector equals the electric field seen by a detector at the same location {\em minus} the electric field seen by a mirror image detector at $\tilde {\bf r} = (-x,y,z)$ in free space, i.e.~ without the mirror interface present. More concretely, the electric field observable ${\mathbf E}_{\text{mirr}}(\mathbf r)$ equals
\begin{eqnarray}\label{methodofimagedetectorsperfectmirror}
{\mathbf E}_{\text{mirr}}(\mathbf r) &=&  {1 \over \eta_b} \, \left[ {\bf E}^{(b)}_{\text{med}}({\mathbf {r}}) - \widetilde {\bf E}^{(b)}_{\text{med}}(\tilde{{\mathbf {r}}}) \right] \Theta (-x) +  {1 \over \eta_a} \, \left[ {\bf E}^{(a)}_{\rm air}(\mathbf r) - \widetilde {\bf E}^{(a)}_{\rm air} (\tilde{\mathbf r}) \right] \Theta(x) \, ,
\end{eqnarray}
if we assume that the $a$ and the $b$ photons evolve as they would in air. Here $\Theta (x)$ denotes the Heaviside step function  
\begin{eqnarray}
\Theta (x) &=& \left\{ \begin{array}{cc} 
1 &  \quad \text{for} \quad  x\geq 0 \, , \\
0 & \quad \text{for} \quad x<0 \\
\end{array} \right.
\label{heaviside}
\end{eqnarray}
and the tilde indicates that a minus sign has been added to the $x$ component of the respective vector. Moreover, $\eta_a$ and $\eta_b$ are normalisation constants, ${\bf E}^{(a)}_{\rm air}(\mathbf r) $ can be found in Eq.~(\ref{EMfieldfree3d}) and $ {\bf E}^{(b)}_{\text{med}}({\mathbf {r}})$ can be obtained from Eq.~ (\ref{quantisedfieldmed3D}) by replacing the $a_{\mathbf k \lambda}$ operators in this equation with $b_{\mathbf k \lambda}$. Notice that the right hand side of Eq.~(\ref{methodofimagedetectorsperfectmirror}) is a superposition of operators whose expectation values evolve as predicted by Maxwell's equations in a dielectric medium and in air, respectively. Hence the observable ${\mathbf E}_{\text{mirr}}(\mathbf r)$ is automatically consistent with Maxwell's equations on both sides of the mirror interface, independent of what values we assign later on to $\eta_a$ and $\eta_b$.

 As mentioned already above, the constants $\eta_{a}$ and $\eta_{b}$ in Eq.~(\ref{methodofimagedetectorsperfectmirror}) are normalisation factors. In the next section, we determine them by demanding that the spontaneous decay rate of an atom $\Gamma_{\text{mirr}}(x)$ in the presence of the mirror surface simplifies  for large atom-mirror distances $|x|$ to $\Gamma_{\rm air}$ or to $\Gamma_{\text{med}}$, respectively,
\begin{eqnarray}\label{decayratexxx} 
\Gamma_{\text{mirr}}(x) &=& \left\{ \begin{array}{cl} \Gamma_{\rm med} & {\rm for} ~ x \to - \infty \, , \\ \Gamma_{\rm air} & {\rm for} ~ x \to \infty \, . \end{array} \right. 
\end{eqnarray}
As we shall see below, doing so we find that $\eta_a = \eta_b = \sqrt{2}$ for highly reflecting mirrors. Interpreting this result is not straightforward. As pointed out already in Ref.~\cite{furtak}, for the experimental setup shown in Fig.~\ref{figpaperlogo}, the mirror Hamiltonian $H_{\rm mirr}$ in Eq.~(\ref{systemhamiltonian2}) does not coincide with the observable for the energy of the quantised EM field left and right from the mirror interface. The expectation values of the former are in general  larger than the expectation values of the latter. Some of the energy of the system is stored inside the mirror interface which makes it difficult to normalise the electric field observable in Eq.~(\ref{methodofimagedetectorsperfectmirror}) correctly.

\subsection{Mirror-coated dielectric media} \label{subsecmethodofimagessemitransparentmirror}

To obtain the electric field observable ${\mathbf E}_{\text{mirr}}(\mathbf r)$ in the presence of a two-sided semi-transparent mirror, we need to superimpose the electric field observables of the corresponding free-space scenarios such that any incoming wave packets evolve eventually into superpositions of reflected and transmitted wave packets with their amplitudes accordingly re-scaled. Taking this into account and generalising Eq.~(\ref{methodofimagedetectorsperfectmirror}) as described in Ref.~\cite{furtak}, we find that
\begin{eqnarray}\label{Efieldinterface}
{\mathbf E}_{\text{mirr}}(\mathbf r) &=& \left[ {1 \over \eta_b} \, {\bf E}^{(b)}_{\text{med}}({\mathbf {r}}) + {r_b \over \eta_b} \, \widetilde {\bf E}^{(b)}_{\text{med}}(\tilde{{\mathbf {r}}}, \phi _1) + {t_a \over \eta_a} \, {\bf E}^{(a)}_{\text{med}}({\mathbf {r}}, \phi_2) \right] \Theta (-x) \nonumber \\
&& + \left[ {1 \over \eta_a} \, {\bf E}^{(a)}_{\rm air}(\mathbf r) + {r_a \over \eta_a} \, \widetilde {\bf E}^{(a)}_{\rm air} (\tilde{\mathbf r}, \phi _3) + {t_b \over \eta_b} \, {\bf E}^{(b)}_{\rm air}({\mathbf r}, \phi_4) \right] \Theta(x) \, . 
\end{eqnarray}
As before, the superscripts $(a)$ and $(b)$ are used here to distinguish light originating from the left and from the right hand side of the mirror interface, respectively. At $t=0$, only the first and the fourth terms in Eq.~(\ref{Efieldinterface}) contribute to the electric field observable ${\mathbf E}_{\text{mirr}}(\mathbf r) $. The remaining terms in Eq.~(\ref{Efieldinterface}) describe the electric field contributions of wave packets which have either been reflected by or transmitted through the mirror interface. The factors in front of those terms are the relevant reflection and transmission rates. Finally, phases $\phi_i$ have been added to describe the phase shifts that the complex electric field amplitudes experience when in contact with the mirror interface. These additional parameters depend on the physical properties of the mirror coating in Fig.~\ref{figpaperlogo}.

In the absence of absorption, energy conservation implies $r_a = r_b$. Moreover the phases $\phi_i$ have to obey certain conditions \cite{phase1,zeilinger}. However, in the presence of coherent light absorption within the mirror surface, $r_a$ and $r_b$ are in general not the same and the phases and rates in Eq.~(\ref{Efieldinterface}) can assume a wide range of different values \cite{Monzon,Jeffers,Pinkse}. Suppose all light approaching the reflecting layer of the mirror interface from the left is absorbed, while light approaching from the right reaches the reflecting layer and some of it is turned around. In this case, we have $r_b =0$, while $r_a \neq 0$. Since absorption is uncontrolled in many practical situations, reflection rates $r_a$ and $r_b$ are in general not the same.

\section{Atomic decay rates in the presence of a mirror interface} \label{sec4}

In this section, we finally determine the normalisation constants $\eta_a$ and $\eta_b$ in Eq.~(\ref{Efieldinterface}) by deriving the spontaneous decay rate $\Gamma_{\rm mirr}(x)$ of a two-level atom in the presence of a mirror-coated dielectric interface as a function of the atom-mirror distance $|x|$. We then demand that this rate simplifies to its well-known free spaces value for large atom-mirror distances (cf.~Eq.~(\ref{decayratexxx})). 

\subsection{Derivation}

As usual in quantum optics, we describe the dynamics of a two-level atom with ground state $|1 \rangle$ and excited state $|2 \rangle$ by a master equation in Lindblad form \cite{stokes}. In the absence of any external interactions, like laser excitation, and in the interaction picture with respect to the free energy of the atom, its density matrix $\rho_{\text{I}}(t) $ is known to evolve according to the differential equation 
\begin{eqnarray}
\dot \rho _{\text{I}}(t) &=& - {1 \over 2} \Gamma_{\rm mirr}(x) \, \left[ \sigma^+ \sigma^- \, \rho_{\text{I}}(t) + \rho_{\text{I}}(t) \, \sigma^+ \sigma^- \right] + \Gamma_{\rm mirr}(x)  \, \sigma^- \, \rho_{\text{I}}(t) \sigma^+ 
\label{masterequation2}
\end{eqnarray}
with $\sigma^- = \ket{2}\bra{1}$ and $\sigma ^+ = \ket{1}\bra{2}$. The last term in this equation equals \cite{furtak,stokes}
\begin{eqnarray}\label{hcond}
\Gamma_{\rm mirr}(x) \, \sigma^- \, \rho_{\text{I}}(t) \sigma^+ \, \Delta t &=&\frac{1}{\hbar^2} \int^{t+\Delta t}_t {\rm d}t' \int ^{t + \Delta t}_t {\rm d}t'' \, \text{Tr}_{\text{mirr}}[ H_\text{I}(t')\ket{0} \rho_{\text{I}}(t)\bra{0}H_\text{I}(t'')] ~~~~
\end{eqnarray}
up to terms in second order in $\Delta t$. Here $\Delta t$ denotes a relatively short time interval $(\Delta t \ll 1/\Gamma_{\rm mirr}(x))$ and $H_{\rm I}(t)$ is the  Hamiltonian of the atom-field system in the interaction picture.

For example, for an atomic dipole inside a dielectric medium with refractive index $n$, the above interaction Hamiltonian $H_{\rm I}(t)$ equals \cite{stokes}
\begin{eqnarray}\label{mediumcoordhint}
H_{\text I}(t) &=& \frac{{\rm i}e}{4 \pi } \sum _{\lambda = 1,2} \int_{\mathbb{R}^3} {\rm d}^3 \mathbf k \, \sqrt{\frac{n^3 \hbar \omega}{\pi \varepsilon}} \, {\rm e}^{{\rm i} n \mathbf k \cdot \mathbf {r}}\, {\rm e}^{{\rm i}(\omega -\omega_0)t} \, \mathbf d^*_{12} \cdot \hat{\mathbf e}_{\mathbf k \lambda} \, \sigma^+\,  a_{\mathbf k \lambda} + \text{H.c.} 
\end{eqnarray}
in the usual dipole and rotating wave approximations and with respect to the free energy of the atom and the quantised EM field near the mirror interface. Here $e$ is the charge of a single electron, ${\bf d}_{12}$ denotes the complex atomic dipole moment and $\omega = \|{\bf k}\|/c$. Moreover, $\hbar \omega_0$ is the energy difference between the ground and the excited state of the atom. Substituting Eq.~(\ref{mediumcoordhint}) into Eq.~(\ref{hcond}), proceeding as usual \cite{furtak,stokes} and evaluating the above integrals, we find that the spontaneous decay rate $\Gamma = \Gamma_{\rm med}$ of an atom inside a dielectric medium equals  
\begin{eqnarray}\label{hcond6}
\Gamma_{\rm med} &=&\frac{n^3 e^2 \omega_0^3 \, \| \mathbf{d}_{12} \|^2}{3 \pi \hbar \varepsilon c_0^3} =  \frac{e^2 \omega^3_0 \|\mathbf d_{12}\|^2 }{3 \pi \varepsilon c^3\hbar} \, .
\end{eqnarray}
For $n=1$ and $\varepsilon = \varepsilon_0$, $\Gamma_{\rm med}$ simplifies to the free space decay rate $\Gamma_{\rm air}$ of an atomic dipole in air,
\begin{eqnarray}
\Gamma_{\text{air}} &=& \frac{e^2 \omega^3_0 \|\mathbf d_{12}\|^2 }{3 \pi \varepsilon_0 c_0^3\hbar} \, .
\label{decayratemedcoords2air}
\end{eqnarray}
It must be noted that in most dielectric media, $\mu$ and $\mu_0$ are very similar \cite{griffiths}. Assuming that $\mu = \mu_0$ and combining the definition of the speed of light in air and in a medium with Eq.~(\ref{defn2}), we obtain Eq.~(\ref{1}). In this case, $\Gamma_{\text{med}}$ and $\Gamma_{\text{air}}$ differ only by a factor $n$ \cite{scheel}. 

To derive the interaction Hamiltonian $H_{\rm I}(t)$ for the experimental setup shown in Fig.~\ref{figpaperlogo}, we notice that it consists of a dielectric medium with mirror coating and an atom at a position ${\bf r}$ in front of the interface. Hence, in the Schr\"odinger picture, its Hamiltonian is of the form
\begin{eqnarray} \label{total}
H_{\text{total}} &=& H_{\text{atom}} + H_\text{mirr} + H_\text{int} \, .
\end{eqnarray}
Here $H_{\text{atom}} = \hbar \omega_0 \, \sigma^+ \sigma^-$ describes the energy of the atom and $H_\text{mirr}$ denotes the energy of the EM field in the presence of an optical interface which can be found in Eq.~(\ref{systemhamiltonian2}). Moreover, $H_\text{int}$ describes the atom-field interaction and equals 
$H_\text{int} = e \, \mathbf d \cdot \mathbf E _{\text{mirr}} (\mathbf r)$ in the usual dipole approximation \cite{stokes}. Here $E _{\text{mirr}} (\mathbf r)$ equals the electric field observable in Eq.~(\ref{Efieldinterface}) at the position ${\bf r}$ of the atom and $\mathbf d = \mathbf d_{12} \, \sigma ^- + \mathbf d_{12} ^* \, \sigma ^+ $ with $\mathbf d_{12} =  \|\mathbf{d}_{12}\| \, (d_1, d_2, d_3)^{\rm T}$ denoting the complex atomic dipole moment with $|d_1|^2 + |d_2|^2 + |d_3|^2 = 1$. Transforming $H_{\rm total}$ into the interaction picture  with respect to the free Hamiltonian $H_0 = H_{\text{atom}} + H_\text{mirr}$ yields the interaction Hamiltonian $H_\text{I}(t) =U^\dagger_0(t,0) \, H_\text{int} \, U_0(t,0)$. Combining Eqs.~(\ref{EMfieldfree3d}) and (\ref{Efieldinterface}) and applying the rotating wave approximation, one can show that this Hamiltonian equals
\begin{eqnarray}\label{interfaceinteractionh}
H_\text{I}(t) &=& \frac{{\rm i}e}{4\pi} \sum _{\lambda = 1,2} \int_{\mathbb R ^3} {\rm d}^3 \mathbf k \, \sqrt{\frac{\hbar \omega}{\pi \varepsilon_0}} \, {\rm e}^{-{\rm i} (\omega - \omega _0)t} \, \left[ {1 \over \eta_a} \, \mathbf{d}_{12}^* \, {\rm e}^{{\rm i} \mathbf k \cdot {\mathbf{r}}} \, a_{\mathbf k \lambda} - \frac{r_a}{\eta_a} \, \widetilde{\mathbf{d}}_{12}^* \, {\rm e}^{{\rm i} \mathbf k \cdot  \tilde{\mathbf {r}}} \, {\rm e}^{{\rm i}\phi_3} \, a_{\mathbf k \lambda} \right. \nonumber\\ 
&& \left. + \frac{t_b}{\eta_b} \, \mathbf{d}_{12}^* \, {\rm e}^{{\rm i} \mathbf k \cdot \mathbf{r}} \, {\rm e}^{{\rm i} \phi_4} \, b_{\mathbf k \lambda}  \right] \cdot \hat{\mathbf e}_{\mathbf{k} \lambda} \, \sigma ^+ + \text{H.c.}
\end{eqnarray}
for an atomic dipole in front of a mirror-coated dielectric medium (cf.~Fig.~\ref{figpaperlogo}).

To calculate its spontaneous decay rate $\Gamma_{\rm mirr}(x)$, we substitute Eq.~(\ref{interfaceinteractionh}) into the right hand side of Eq.~(\ref{hcond}). Doing so one can show that
\begin{eqnarray}\label{hcond7}
\Gamma_{\rm mirr}(x) &=&\frac{e^2}{16 \pi^3 \hbar \varepsilon_0 \, \Delta t} \int^{t+\Delta t}_t {\rm d}t' \int ^{t + \Delta t}_t {\rm d}t'' \int_{\mathbb{R}^3} {\rm d}^3 \mathbf k \nonumber \\ 
&& \times \sum _{\lambda = 1,2}  \omega \, \bigg[ {1 \over \eta_a^2} \left| \mathbf{d}_{12}^* \cdot \hat{\mathbf e}_{\mathbf{k} \lambda} \, {\rm e}^{{\rm i} \mathbf k \cdot \mathbf {r}} - r_a \, \widetilde{\mathbf{d}}^*_{12} \cdot \hat{\mathbf e}_{\mathbf{k} \lambda} \, {\rm e}^{{\rm i} \mathbf k \cdot  \tilde{\mathbf {r}}} \, {\rm e}^{{\rm i} \phi_3} \right|^2 + {t_b^2 \over \eta_{b}^2} \, | \mathbf{d}_{12} \cdot \hat{\mathbf e}_{\mathbf{k} \lambda}|^2\bigg] \, {\rm e}^{{\rm i}(\omega -\omega_0)(t'-t'')} \, . \nonumber \\
\end{eqnarray}
Before performing any time integrations, we substitute $s'=t'- t$ and $s''=t''- t$ and notice that the time integrals
\begin{eqnarray}\label{hcond3}
\int^{t+\Delta t}_t {\rm d}t' \int ^{t + \Delta t}_t {\rm d}t'' \, {\rm e}^{{\rm i}(\omega -\omega_0)(t'-t'')}
&=& \int_0^{\Delta t} {\rm d}s' \int_{0}^{\Delta t} {\rm d}s'' \, {\rm e}^{{\rm i}(\omega -\omega_0)(s'-s'')} \nonumber \\
&=& 2 {\rm Re} \left[ \int_0^{\Delta t} {\rm d}s' \int_{0}^{s'} {\rm d}s'' \, {\rm e}^{{\rm i}(\omega -\omega_0)(s'-s'')} \right] ~~~
\end{eqnarray}
are independent of $t$ and always real. Moreover we know that $\Delta t$ and therefore also almost all $s'$ are much larger than $1/\omega_0$. Hence we can safely assume that 
\begin{eqnarray}\label{hcond4}
\int_{0}^{s'} {\rm d}s'' \, {\rm e}^{-{\rm i}(\omega -\omega_0)s''} = \int_{0}^\infty {\rm d}s'' \, {\rm e}^{-{\rm i}(\omega -\omega_0) s''} 
= \pi \, \delta(\omega -\omega_0) 
\end{eqnarray}
up to an imaginary part which does not contribute to later integrals. To perform the remaining $\mathbf k$ integration we use polar coordinates and introduce the vectors
\begin{eqnarray} \label{vectors}
\mathbf k = k
\begin{pmatrix}
\cos\vartheta\\
\cos\varphi\sin\vartheta\\
\sin\varphi\sin\vartheta
\end{pmatrix}, ~~
\hat{\mathbf e}_{\mathbf{k} 1} = 
\begin{pmatrix}
0 \\
\sin\varphi \\
- \cos\varphi
\end{pmatrix} , ~~ 
\hat{\mathbf e}_{\mathbf{k} 2} = 
\begin{pmatrix}
\sin\vartheta\\
- \cos\varphi\cos\vartheta\\
- \sin\varphi\cos\vartheta
\end{pmatrix} 
\end{eqnarray}
with $\omega = c_0k$, resulting in
\begin{eqnarray}
\int_{\mathbb R ^3} d^3 \mathbf k =  \int_{0}^{\infty} d\omega \int_0^\pi d\vartheta \int_0^{2\pi}d\varphi \, \frac{\omega ^2}{c_0^3} \, \sin\vartheta \, .
\label{intkredefine}
\end{eqnarray}
Using the above equations and performing time and frequency integrations, while denoting the atom-mirror distance by $x$ such that ${\bf r} - \tilde{\bf r} = 2x$, on can now show that 
\begin{eqnarray}\label{hcond8}
\Gamma_{\rm mirr}(x) &=& \frac{e^2 \omega_0^3 \, \| \mathbf{d}_{12} \|^2}{8 \pi^2 \hbar \varepsilon_0 c_0^3} \int_0^\pi {\rm d}\vartheta \int_0^{2\pi} {\rm d}\varphi \, \sin \vartheta \nonumber \\
&& \hspace*{-2cm} \times \bigg[ \left({1 \over \eta_a^2} + { t_b^2 \over \eta_b^2}\right) \left( \left| d_2 \sin\varphi - d_3 \cos\varphi \right|^2 + \left| d_1 \, \sin\vartheta - d_2 \, \cos\varphi \cos\vartheta - d_3 \, \sin\varphi \cos\vartheta \right|^2 \right) \nonumber \\
&& \hspace*{-2cm} + {r_a^2 \over \eta_a^2} \left( \left| d_2 \sin\varphi - d_3 \cos\varphi \right|^2 + \left| d_1 \, \sin\vartheta + d_2 \, \cos\varphi \cos\vartheta + d_3 \, \sin\varphi \cos\vartheta \right|^2 \right) \nonumber \\
&& \hspace*{-2cm} + {r_a \over \eta_{a}^2} \Big( \left( d_1^* \sin \vartheta - d_2^* \cos \varphi \cos \vartheta - d_3^* \sin \varphi \cos \vartheta \right) \left( d_1 \sin \vartheta + d_2 \cos \varphi \cos \vartheta + d_3 \sin \varphi \cos \vartheta \right) \nonumber \\
&& \hspace*{-2cm} - \left| d_2 \sin \varphi - d_3 \cos \varphi \right|^2 \Big) \,  {\rm e}^{2{\rm i} k_0 x \cos \vartheta} \, {\rm e}^{-{\rm i} \phi_3} + {\rm c.c.} \bigg] 
\end{eqnarray}
with $k_0=\omega_0/c_0$. Next we perform the $\varphi$ integration, substitute $u = \cos\vartheta$ and use the relation $|d_2|^2 + |d_3|^2 = 1 - |d_1|^2 $ to obtain the integral
\begin{eqnarray}\label{hcond9}
\Gamma_{\rm mirr}(x) &=& \frac{e^2 \omega_0^3 \, \| \mathbf{d}_{12} \|^2}{8 \pi \hbar \varepsilon_0 c_0^3} \int_{-1}^1 {\rm d}u \bigg[ 
\left({1 + r_a^2 \over \eta_a^2}+ { t_b^2 \over \eta_b^2}\right) \left( 1 + |d_1|^2 + \left(1- 3 |d_1|^2\right) u^2 \right) \nonumber \\
&& - {2r_a \over \eta_{a}^2} \left( 1 - 3 |d_1|^2 + \left(1+ |d_1|^2 \right) u^2 \right) \cos(2k_0x u - \phi_3) \bigg] 
\end{eqnarray}
with $\cos(2k_0x u - \phi_3) = \cos(2k_0x u ) \cos(\phi_3) + \sin(2k_0x u ) \sin( \phi_3)$. Finally also performing the $u$ integration in Eq.~(\ref{hcond9}), we obtain the spontaneous decay rate  
\begin{eqnarray}\label{decayrate} 
{\Gamma_{\text{mirr}}(x) \over \Gamma_{\rm air}} &=& \frac{1 + r_a^2}{\eta^2_a} + \frac{t_b^2}{\eta^2_b} + \frac{3r_a}{\eta^2_a} \, \cos(\phi_3) \left[ \big(1- |d_1|^2\big) \frac{\sin(2k_0 x)}{2k_0 x} \right. \nonumber\\ 
&& \left. + \big(1+|d_1|^2\big) \left( \frac{\cos(2k_0 x)}{(2k_0 x)^2} - \frac{\sin(2k_0 x)}{(2k_0 x)^3}\right) \right] ~~~~
\end{eqnarray}
for $x> 0$. Here $|d_1|^2$ denotes the relative overlap of the normalised atomic dipole moment vector $\mathbf d_{12}/\| \mathbf d_{12} \|$ with the $x$ axis. For $|d_1|^2 = 0$, the atomic dipole aligns parallel to the mirror interface, while it aligns in a perpendicular fashion when $|d_1|^2 = 1$. An equivalent expression for $\Gamma_{\text{mirr}}(x) $ can be derived for the case $x<0$. The result is the same as in Eq.~(\ref{decayrate}) but with the subscripts $a$ and $b$ interchanged and with $\Gamma_{\rm air}$ and $\phi_3$ replaced by $\Gamma_{\rm med}$ and $\phi_1$, respectively. The above calculations are well justified, as long as the atom-mirror distance $|x|$ is not too large such that the travel time of light between the atom and the mirror surface remains negligible \cite{zoller}. 

The only other simplification which has been made in the derivation of Eq.~(\ref{decayrate}) is the negligence of surface plasmons and evanescent modes. These modes can provide an additional decay channel for atomic excitation and their presence can lead to an increase of emission rates. However, here we assume that $x$ should be large enough for interactions with surface plasmons and evanescent modes not to become important.

\subsection{The normalisation constants $\eta_a$ and $\eta_b$} \label{etas}

However, before we can make more quantitative predictions, we need to determine the normalisation factors $\eta_a$ and $\eta_b$. To do so, we demand that the spontaneous decay rate $\Gamma_{\text{mirr}}(x)$ in Eq.~(\ref{decayrate}) simplifies to the expressions in Eqs.~(\ref{hcond6}) and (\ref{decayratemedcoords2air}), respectively, for large atom-mirror distances $|x|$, as suggested in Eq.~(\ref{decayratexxx}). It is relatively straightforward to show that this applies when 
\begin{eqnarray} \label{oma}
\frac{1+r_a^2}{\eta_a^2} + \frac{t_b^2}{\eta_b^2} = \frac{1 + r_b^2}{\eta^2_b} + \frac{t_a^2}{\eta^2_a} &=& 1 
\end{eqnarray}
which applies when
\begin{eqnarray}
\eta^2_{a} &=& 1+ r_a^2 + {1+ r_a^2 - t_a^2 \over 1 + r_b^2-t_b^2} \, t_b^2 \, , \nonumber \\
\eta^2_{b} &=& 1+ r_b^2 + {1+ r_b^2 - t_b^2 \over 1 + r_a^2-t_a^2} \, t_a^2   \, .
\label{normalisationconsts}
\end{eqnarray}
Both normalisation factors $\eta_a^2$ and $\eta_b^2$ are always larger than one. They only equal one, in the absence of the mirror interface, i.e.~when all reflection and transmission rates are equal to zero. In this case, the electric field observable in Eq.~(\ref{Efieldinterface}) simplifies to its free space value.

\begin{figure}[t]
	\centering
	\includegraphics[width=17.2cm]{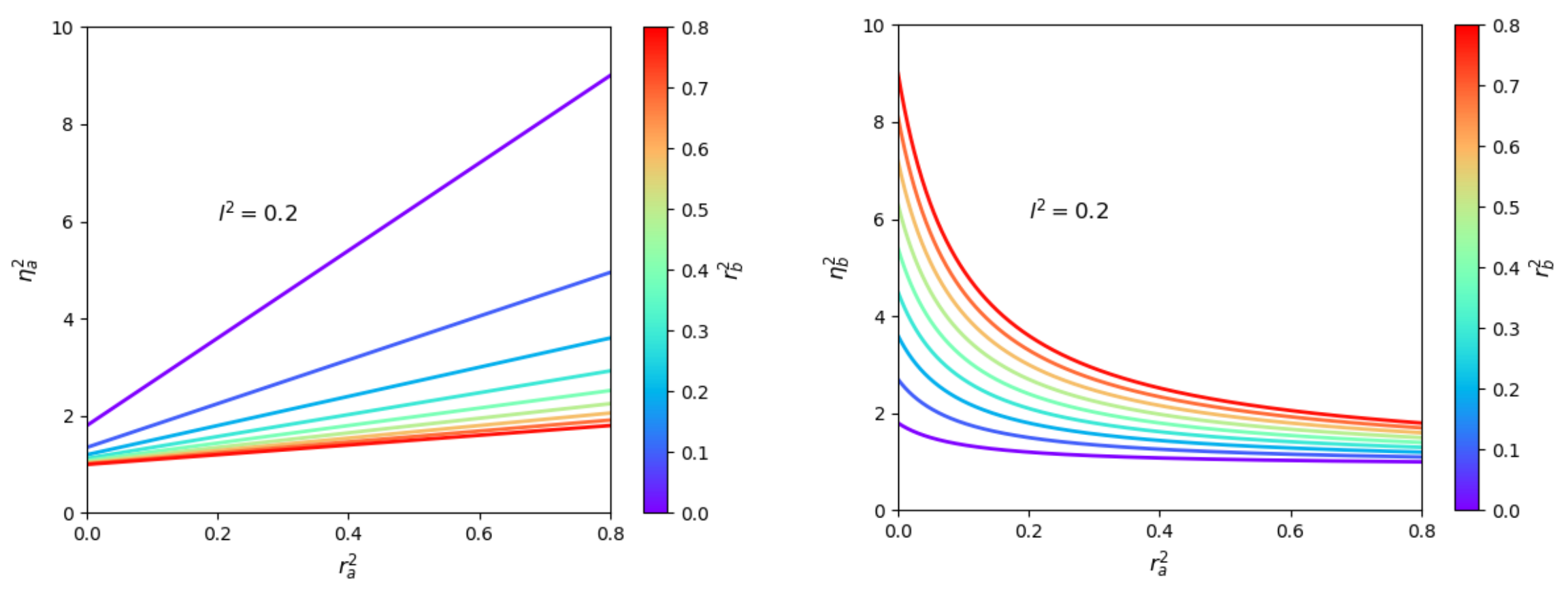}
	\caption{ The normalisation factors $\eta_a^2$ and $\eta_b^2$ of the electric field observable ${\bf E}_{\rm mirr}({\bf r})$ in Eq.~(\ref{Efieldinterface}) as a function of the reflection rates $r_a$ and $r_b$ for mirror loss rates $l_a=l_b = l$ with $l^2=0.2$. In the absence of loss, $\eta_a^2$ and $\eta_b^2$ only vary between 1 and 2. However, in the presence of light absorption within the mirror interface, these rates can assume much larger values, while still being bound from below by 1.}
	\label{figetafunction}
\end{figure}

For {\em symmetric mirrors}, we have $r_a = r_b = r$ and $t_a = t_b = t$. Substituting these constants into the above expressions, they simplify and we find that $\eta_a^2=\eta_b^2= \eta^2$ with
\begin{eqnarray}
\eta^2 &=& 1 + r^2 + t^2 
\end{eqnarray}
which can assume and value between 1 and 2. For example, for highly-reflecting symmetric mirrors with $r=1$ and $t=0$ we have $\eta^2 = 2$ \cite{furtak}. However, for {\em asymmetric mirrors}, $\eta_a$ and $\eta_b$ are no longer bound from above. This is illustrated in Fig.~\ref{figetafunction} which shows $\eta_a^2$ and $\eta_b^2$ for an asymmetric mirror with equal loss rates $l_a = l_b = l$ and $l^2 = 0.2$. These rates are defined such that energy is conserved \cite{Monzon,Jeffers,Pinkse} and
\begin{eqnarray}
r_a^2 + t_a^2 + l_a^2 ~=~ r_b^2 + t_b^2 + l_b^2 &=& 1 \, .
\end{eqnarray}
For example, suppose the reflection rate $r_a$ is relatively large, while $r_b$ is very small, as it applies when the mirror surface is very smooth and highly reflective on the right hand side but rough and highly dispersive on the left (cf.~Fig.~\ref{figpaperlogo}). In this case, $\eta_a$ can be significantly larger than $\eta_b$, if the loss rates $l_a$ and $l_b$ are similar in size (cf.~Fig.~\ref{figetafunction}). This implies that the electric field observable ${\mathbf E}_{\text{mirr}}(\mathbf r)$ in Eq.~(\ref{Efieldinterface}) is dominated by the contributions of the $b$ rather than the $a$ photons. This can be understood by taking into account that the $b$ photons are present on both sides of the mirror interface in this case while, to a very good approximation, the $a$ photons can only be seen on one side.

\subsection{Discussion}

\begin{figure}[t]
	\centering
        \includegraphics[width=17.2cm]{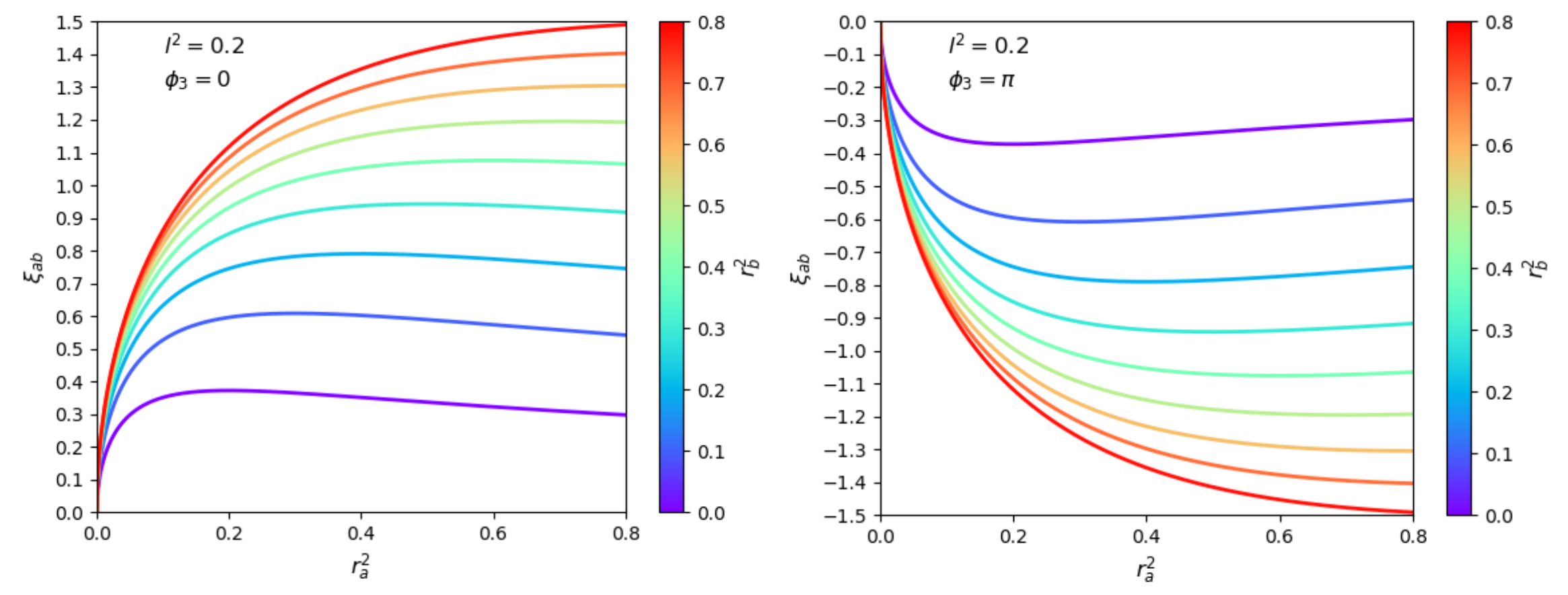}
	\caption{ The mirror parameter $\xi_{ab}$ in Eq.~(\ref{xi}) as a function of the reflection rates $r_a$ and $r_b$ of the mirror interface for $\phi_3=0$ and $\phi_3=\pi$. As in Fig.~\ref{figetafunction}, we consider non-zero absorption rates and assume $l_a=l_b=l$ with $l^2=0.2$. In general, the mirror parameter $\xi_{ab}$ can vary between $-1.5$ and $1.5$.}
	\label{figasymmetricalmirrorenhancement}
\end{figure}

In this subsection, we have a closer look at the spontaneous decay rates $\Gamma_{\text{mirr}}(x) $ of an atom on the right hand side of a mirror-coated interface with coherent light absorption where $x$ is positive. Using Eq.~(\ref{oma}), $\Gamma_{\text{mirr}}(x) $ in Eq.~(\ref{decayrate}) simplifies to 
\begin{eqnarray}\label{decayratesimplified} 
{\Gamma_{\text{mirr}}(x) \over \Gamma_{\rm air}} &=&1 + \xi_{ab} \left[ \big(1-|d_1|^2 \big) \frac{\sin(2k_0 x)}{2k_0 x} + \big(1+|d_1|^2 \big) \left( \frac{\cos(2k_0 x)}{(2k_0 x)^2} - \frac{\sin(2k_0 x)}{(2k_0 x)^3} \right) \right] ~~
\end{eqnarray}
with the mirror parameter $\xi_{ab}$ given by  
\begin{eqnarray}\label{xi} 
\xi_{ab} &=& \frac{3 r_a }{\eta^2_a} \, \cos (\phi_3)  \, . 
\end{eqnarray}
This equation shows that the difference between the spontaneous decay rates $\Gamma_{\text{mirr}}(x)$ and $\Gamma_{\rm air}$ depends on the phase $\phi_3$ which is the phase that complex electric field amplitudes accumulate upon reflection by the mirror surface on the same side as the atom. It also depends on the orientation $|d_1|^2$ of the atomic dipole moment with respect to the mirror surface, as one would intuitively expect.

However, a closer look at Eq.~(\ref{decayratesimplified}) also shows that the spontaneous decay rate $\Gamma_{\text{mirr}}(x)$ depends in addition on {\em all} the reflection and transmission rates of the mirror interface. This might seem surprising but remember that the dipole interaction between the atom and the surrounding free radiation field plays an integral role in the spontaneous emission of a photon (cf.~Eq.~(\ref{total})). In the experimental setup in Fig.~\ref{figpaperlogo}, the atom couples to incoming, reflected and transmitted photon modes which leads to interference effects and the strong dependence of $\Gamma_{\text{mirr}}(x)$ on the atom-mirror distance $x$. Moreover, the strength of the atom-field interaction depends on the magnitude of the electric field observable ${\bf E}_{\rm mirr} ({\bf r})$ at the position {\bf r} of the atom (cf.~Eq.~(\ref{Efieldinterface})). As we have seen in the discussion at the end of Section \ref{subsecmethodofimagesperfectmirror}, normalising this observable is not straightforward, since the total energy of the $a$ and the $b$ photons is shared between the quantised EM field {\em and} the mirror interface \cite{furtak}.

\begin{figure}[t]
	\centering
        \includegraphics[width=17.2cm]{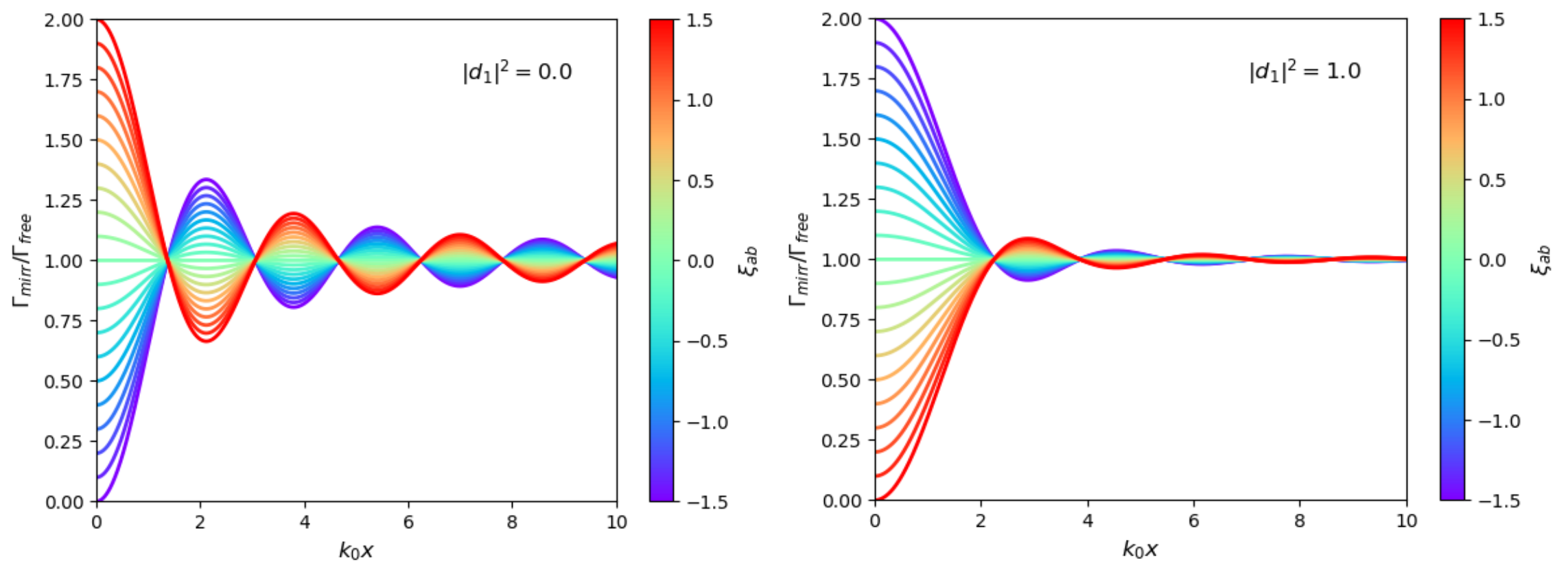}
	\caption{The spontaneous decay rate $\Gamma_{\rm mirr}(x)$ in Eq.~(\ref{decayratesimplified}) as a function of the atom-mirror distance $x$ for different mirror parameters $\xi_{ab}$. When the atomic dipole moment is parallel to the mirror interface ($|d_1|^2 = 0$), the variations of $\Gamma_{\rm mirr}(x)$ are more long-range than in the case of a perpendicular atomic dipole moment $(|d_1|^2 = 1)$. In both cases, the decay rate $\Gamma_{\rm mirr}(x)$ converges and assumes its free space value $\Gamma_{\rm air}$ when the atom-mirror distances $x$ becomes much larger than the wave length of the emitted light.}
	\label{mudep5}
\end{figure}

The reason for the dependence of $\Gamma_{\text{mirr}}(x)$ on $r_a$, $r_b$, $t_a$ and $t_b$ is its dependence on the mirror constant $\xi_{ab}$. Fig.~\ref{figasymmetricalmirrorenhancement} shows that $\xi_{ab}$ can assume any value between $-1.5$ and $1.5$. For example, the case $|\xi_{ab}| = 1.5$ corresponds to a perfectly-reflecting mirror with $r_a = 1$, $t_b=0$ and $\phi_3 = 0$ or $\phi_3 = \pi$. From Fig.~\ref{mudep5} we see that $\Gamma_{\rm mirr}(x)$ can therefore assume any value between $0$ and $2 \Gamma_{\rm air}$. The presence of loss in the mirror interface reduces the amount of light which can be transmitted and changes $\xi_{ab}$ in a relatively complex way (cf.~Eqs.~(\ref{normalisationconsts}) and (\ref{xi})). For example, increasing $l_a$ results in a reduction of $\xi_{ab}$, while increasing $l_b$ results in general in an increase of $\xi_{ab}$. To better illustrate the dependence of $\Gamma_{\rm mirr}(x)$ on mirror parameters, we will now have a closer look at concrete examples. First it will be shown that our approach reproduces well-known results for loss-less symmetric mirrors, thereby verifying the consistency of our approach. Afterwards, we will discuss how the coherent absorption of light in the mirror surface alters atomic decay rates.

\subsubsection{Dielectric media without mirror coatings}

\begin{figure}[t]
	\centering
        \includegraphics[width=17.2cm]{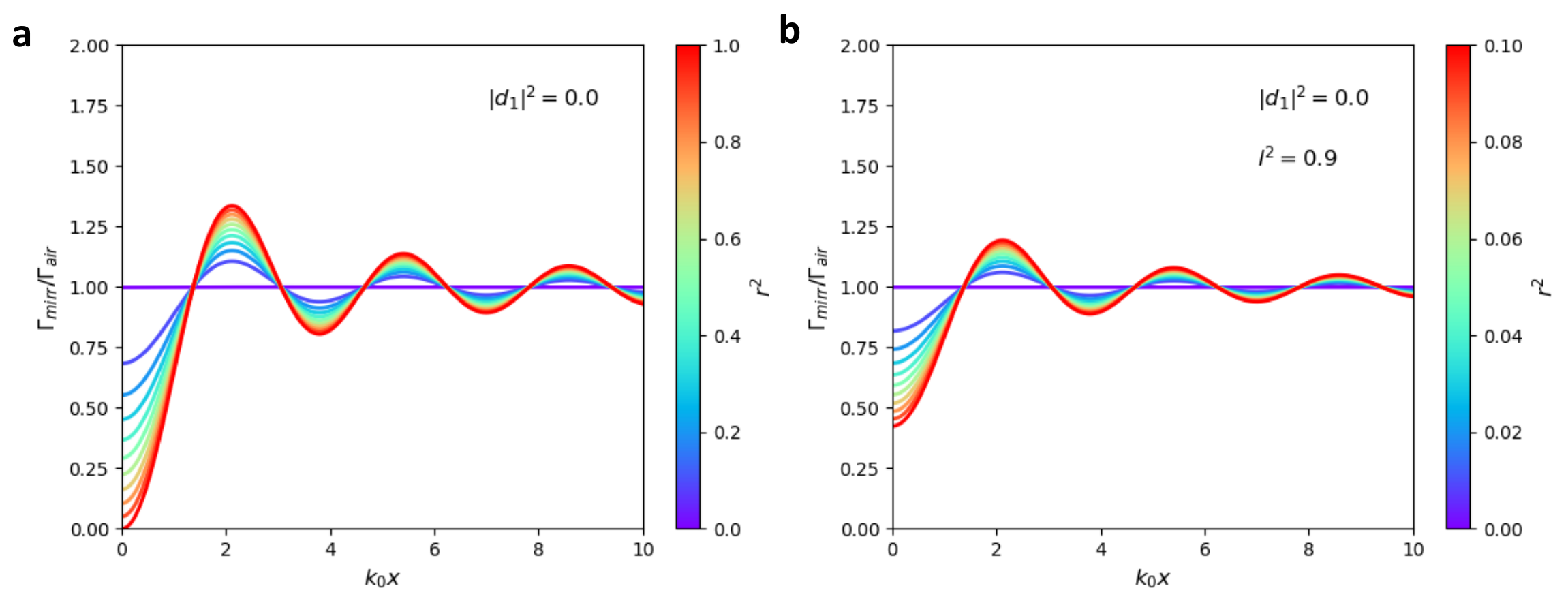}
	\caption{The spontaneous decay rate $\Gamma_{\rm mirr}(x)$ in Eq.~(\ref{decayratesimplified}) of an atom in front of a dielectric medium as a function of the atom-mirror distance $x$ for $\phi_3=\pi$ and $|d_1|^2=0$. ({\bf a}) Here we ignore the mirror-coating and the reflection and transmission rates are chosen as suggested in Eq.~(\ref{oma2}). ({\bf b}) Here the mirror coating is taken into account. Again we assume that $r_a = r_b=r$. However, instead of ignoring the possible absorption of light in the mirror interface and to better showcase its effects we consider relatively high loss rates with $l_a = l_b = l$ and $l^2=0.9$. Nevertheless, ({\bf a}) and ({\bf b}) have many similarities.}
	\label{figmudep} 
\end{figure}

In the absence of any coating, energy is conserved and the overall transition matrix for incoming photons needs to be unitary. Taking this into account one can show that \cite{phase1,zeilinger} 
\begin{eqnarray} \label{oma2}
l_a , \, l_b = 0 \, ; ~~ r_a , \, r_b = r \, ; ~~ t_a , \,  t_b = (1 - r^2)^{1/2} 
\end{eqnarray}
in this case. As a result, the mirror constant $\xi_{ab}$ in Eq.~(\ref{xi}) simplifies to 
\begin{eqnarray}\label{xi2} 
\xi_{ab} &=& \frac{3 r}{2} \, \cos (\phi_3)  \, . 
\end{eqnarray}
and depends only on r and $\phi_3$. Fig.~\ref{figmudep}(a) illustrates the dependence of $\Gamma_{\rm mirr}(x)$ on $r$ and $x$ for two different orientations of the atomic dipole moment $|d_1|^2$ and $\phi_3=\pi$. Because of the dependence of the reflection rate $r$ of a dielectric medium on its refractive index $n$ \cite{Hecht},
\begin{eqnarray}\label{xi2} 
r &=& \frac{n-1}{n+1} \, ,
\end{eqnarray}
the spontaneous decay rate $\Gamma_{\rm mirr}(x)$ depends on the optical properties of the media on both sides of the interface. This observation is in agreement with actual experiments \cite{drexhage,chance,blatt,creatore2}. It is also in agreement with the literature where the spontaneous decay of an atom in the presence of a dielectric medium has already been studied in great detail \cite{carniglia,creatore,khosravi, wylie,snoeks,yeung,rikken,xu,eberlein,wang,falinejad}.

In the special case of a highly-reflecting mirror, which adds a minus sign to the electric field amplitude upon reflection \cite{morawitz,stehl,milonni,arnoldus,drabe,meschede,amos,matloob,zoller,walther}, we have $\phi_3=\pi$, $r = 1$ and $\xi_{ab} = 1.5$. At $x=0$, incoming and reflected light interferes destructively and the resulting $y$ and the $z$ components of the electric field vanish along the mirror surface. If there is no electric field to couple to, then there is no atom-field interaction and the atom cannot decay.  In contrast to this, an atomic dipole which aligns parallel to the mirror surface couples only to the $x$ component of the electric field. This component is now $\sqrt{2}$ times its usual amplitude which results in an enhanced spontaneous decay rate of $\Gamma_{\text{mirr}}(x) = 2 \Gamma_{\rm air}$. A closer look at the $r=1$ case in Fig.~\ref{figmudep}(a) shows that this is indeed the case.

\subsubsection{Dielectric media with mirror coatings}

\begin{figure}[t]
	\centering
	\includegraphics[width=8.6cm]{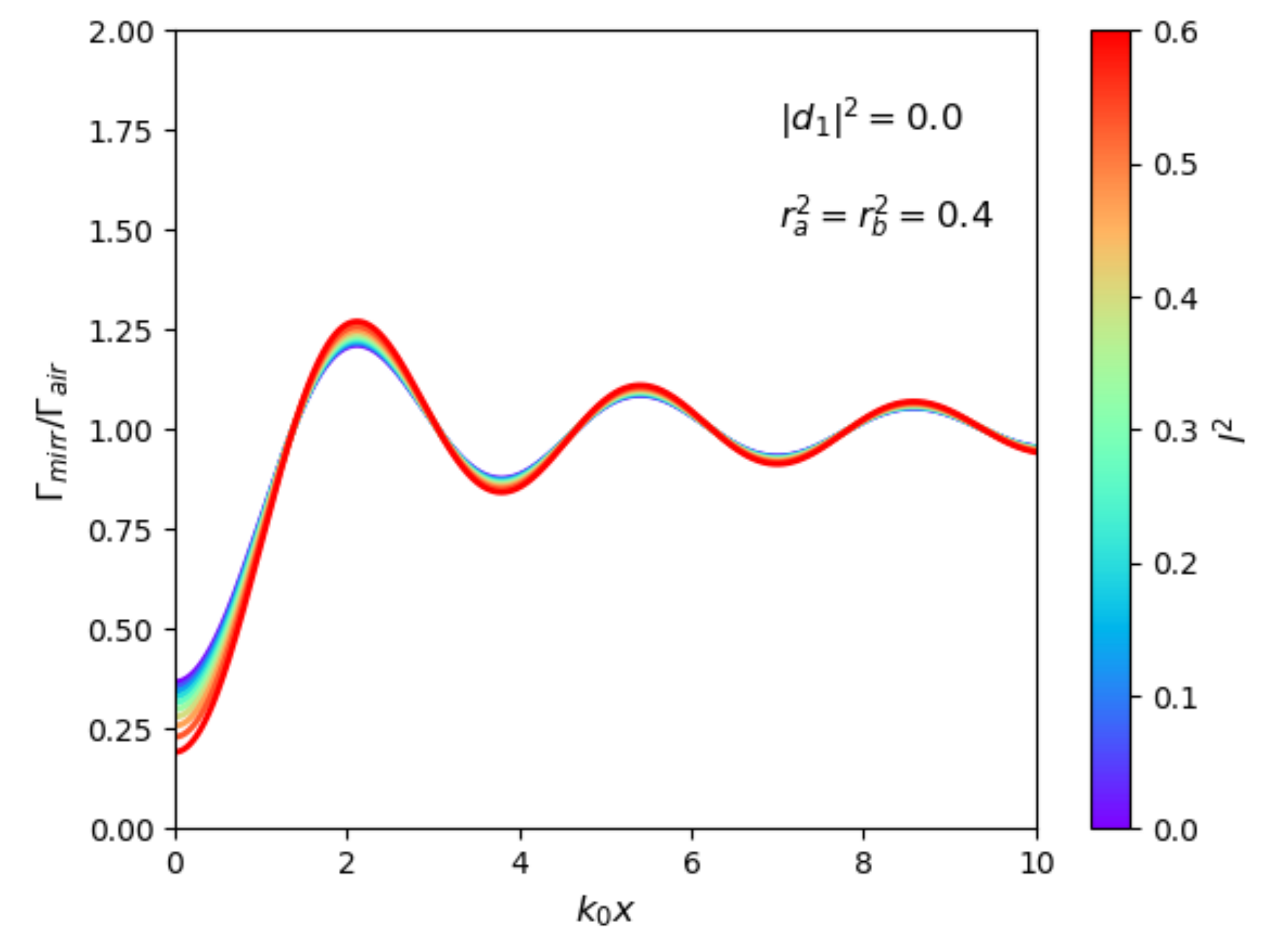}
	\caption{ The spontaneous decay rate $\Gamma_{\rm mirr}(x)$  in Eq.~(\ref{decayratesimplified}) of an atom in front of a mirror-coated dielectric medium as a function of  the atom-mirror distance $x$. Here $r_a = r_b = r$ with $r^2 =0.4$, $\phi_3=\pi$ and $|d_1|^2=0$. Moreover, we assume that the loss rates $l_a$ and $l_b$ are the same and $l^2$ with $l_a=l_b=l$ varies between 0 and $0.6$. Again we find that the possible absorption of light in the mirror interface does not change the atomic decay rate of the atom very much.}
	\label{figldep}
\end{figure}

In the presence of mirror coatings, the possible absorption of light in the interface needs to be taken into account. As we have seen in Section \ref{sec2}, in the quantum mirror image detector method \cite{furtak}, this is done by evolving photon states in exactly the same way as they would evolve in free space, i.e.~without reducing their energy in time. However, as one can see from Eq.~(\ref{Efieldinterface}), photons which have either been transmitted or reflected by the mirror interface contribute less the to the electric field observable ${\bf E}_{\rm mirr}({\bf r})$ at the location of the atom than photons which have not met the mirror. Intuitively, we would expect a much weaker change of the spontaneous decay rate $\Gamma_{\text{mirr}}(x)$ with the atom-mirror distance $x$ than in the absence of losses. Indeed, as the calculations in Section \ref{etas} show, the presence of non-zero loss rates, 
\begin{eqnarray}
l_a, \, l_b &\neq & 0 \, , 
\end{eqnarray}
changes the normalisation constant $\eta_a$ of the electric field observable ${\bf E}({\bf r})$ in Eq.~(\ref{Efieldinterface}), thereby also altering the mirror constant $\xi_{ab}$ in Eq.~(\ref{xi}). However, there are more similarities between $\Gamma_{\text{mirr}}(x)$ with and without losses than what one might naively expect. Fig.~\ref{figmudep}(b) shows $\Gamma_{\rm mirr}(x)$ for a case with a significant amount of light absorption in the mirror interface ($l_a = l_b = l$ with $l^2 = 0.9$).  Nevertheless, Figs.~\ref{figmudep}(a) and \ref{figmudep}(b) both show a strong dependence of $\Gamma_{\rm mirr}(x)$ on the atom-mirror distance $x$. While $\Gamma_{\rm mirr}(x)$ varies between 0 and 2 in one case, it varies between $0.4$ and $1.6$ in the other.

Fig.~\ref{figldep} shows cases, where the reflection rates $r_a = r_b = r$ are fixed and $r^2=0.4$, while the loss rate $l$ changes between 0 and its maximum possible value of $r$. As in Fig.~\ref{figmudep}(b), we observe a relatively weak dependence on the spontaneous decay rate $\Gamma_{\rm mirr}(x)$ on loss rates of the mirror interface. The most significant effect of the absorption of light in the mirror interface is seen for relatively small values of $x$ which matches the results presented for example in Refs.~\cite{yeung,eberlein}.  

\begin{figure}
	\centering
        \includegraphics[width=17.2cm]{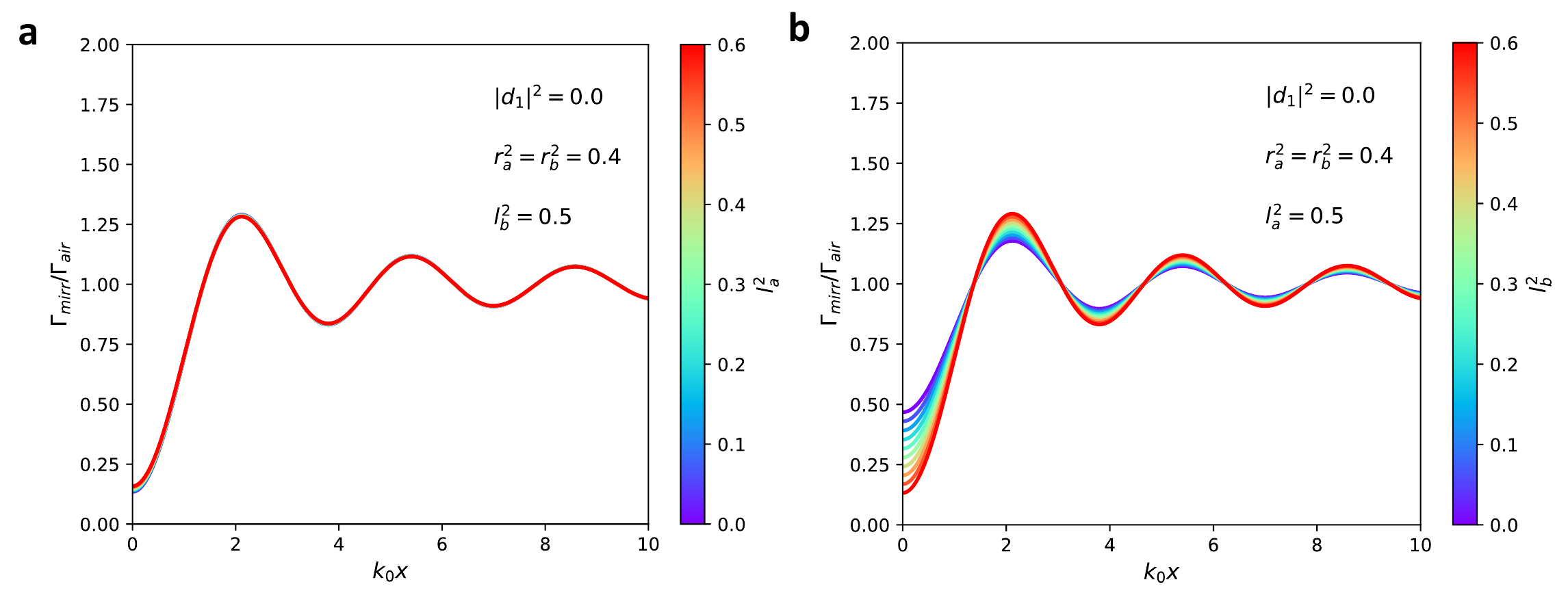}
	\caption{The spontaneous decay rate $\Gamma_{\rm mirr}(x)$  in Eq.~(\ref{decayratesimplified}) of an atom in front of a mirror-coated dielectric medium as a function of  the atom-mirror distance $x$. As in Fig.~\ref{figldep}, $r_a = r_b =r$ with $r^2 = 0.4$, $\phi_3=\pi$ and $|d_1|^2=0$. The figure illustrates that $\Gamma_{\rm mirr}(x)$ depends on the loss rates $l_a$ and $l_b$  in different ways. ({\bf a}) For example, varying $l_a$ (which is the loss rate of light approaching the mirror from the same side as the atom) has almost no effect on $\Gamma_{\rm mirr}(x)$. ({\bf b}) However, changing $l_b$ (which is the loss rate of light approaching the mirror from the opposite side as the atom) changes $\Gamma_{\rm mirr}(x)$ in a much more significant way.}
	\label{figmudep5}
\end{figure}

Finally, Fig.~\ref{figmudep5} shows that the spontaneous decay rate $\Gamma_{\rm mirr}(x)$ depends differently on the loss rates $l_a$ and $l_b$. For example, varying $l_a$ (which is the loss rate of light approaching the mirror from the same side as the atom) while keeping $r_a$ the same has almost no effect on the size of the spontaneous decay rate $\Gamma_{\rm mirr}(x)$ (cf.~Fig.~\ref{figmudep5}(a)). This can be understood by noticing that light which has left the atom no longer affects its dynamics. Once a photon has been emitted, it does not matter whether it is absorbed in the mirror surface, by a far-away detector or by the walls of the laboratory. In contrast to this, changing $l_b$ (which is the loss rate of light approaching the mirror from the opposite side as the atom) can have a noticeable effect on the spontaneous decay rate $\Gamma_{\rm mirr}(x)$ (cf.~Fig.~\ref{figmudep5}(b)). For example, increasing $l_b$ while keeping $r_b$ the same can result in an increase of the dependence of $\Gamma_{\rm mirr}(x)$ on the atom-mirror distance $x$. This occurs due to a reduction of the normalisation constant $\eta_a$ of the electric field observable ${\mathbf E}_{\text{mirr}}(\mathbf r) $ in Eq.~(\ref{Efieldinterface}) which leads to an increase of the dipole interaction between the atom and the $b$ photons.
  
\section{Conclusions} \label{secconclusion}

The fluorescence properties of an atomic dipole depend on the so-called local density of states of the quantised EM field \cite{Bart,Bart2} which itself depends in a complex way on the properties of all of its surroundings. For example, as this paper illustrates, the spontaneous decay rate of an atom near a mirror-coated interface depends on the reflection and transmission rates, $r_a$, $r_b$, $t_a$ and $t_b$, of light approaching the mirror from both sides (cf.~Fig.~\ref{figpaperlogo}). While standard methods, which are based on the calculation of Greens functions or on the introduction of triplet modes (cf.~e.g.~Refs.~\cite{creatore,khosravi,eberlein}), already yield good agreement with experimental findings, this paper aims to provide more physical insight. The potential coherent absorption of light in the interface is explicitly taken into account by assuming that the mirror does not change the shape of incoming wave packets but only reduces amplitudes by given rates.

To obtain an expression for the electric field observable ${\bf E}_{\rm mirr}({\bf r})$ in the presence of a mirror-coated dielectric medium, this paper employs the quantum mirror image detector method \cite{furtak}, doubles the standard Hilbert space of the EM field and maps the dynamics of incoming wave packets onto their dynamics in analogous free space scenarios. In this way, we are able to obtain an expression which is consistent with Maxwell's equations but contains two unknown normalisation factors $\eta_a$ and $\eta_b$ (cf.~Eq.~(\ref{Efieldinterface})). These constants cannot be derived by simply demanding that the energy observable of the EM field and the Hamiltonian of the experimental setup in Fig.~\ref{figpaperlogo} are the same \cite{furtak}. Instead we demand locality and assume that the spontaneous decay rate $\Gamma_{\rm mirr}(x)$ of an atom at a relatively large distance $|x|$ from the mirror interface coincides with its respective free space rates (cf.~Eq.~(\ref{decayratexxx})). 

The main difference between the current paper and earlier work \cite{furtak} is that this paper considers a more general scenario. It is emphasised that the quantum optical properties of the atom depend on the characteristics of the media on both sides of the mirror interface. It is also shown that non-zero loss rates do not necessarily reduce the effect of the mirror by as much as one might naively expect. For example, the spontaneous decay rate of an atom can exhibit a relatively strong dependence on the atom-mirror distance $x$ even for loss rates $l_a^2$ and $l_b^2$ as large as 0.9 (cf.~Fig.~\ref{figmudep}).  In agreement with other authors \cite{yeung,eberlein,chance1975}, we find that the effect of absorption in the medium is most felt by dipole moments close to the interface. 

\section*{Conflict of Interest Statement}

The authors declare that the research was conducted in the absence of any commercial or financial relationships that could be construed as a potential conflict of interest.

\section*{Author Contributions}

All authors contributed to the conception and design of this study. BD and AB wrote the first draft of the manuscript. BD, NF and AB performed and checked the analytical calculations. BD made all the figures in the manuscript with the help of NF. All authors contributed to manuscript revision and read and approved the submitted version.

\section*{Funding}

We acknowledge financial support from the Oxford Quantum Technology Hub NQIT (grant number EP/M013243/1).  



\section*{Data Availability Statement}
Statement of compliance with EPSRC policy framework on research data: This publication is theoretical work that does not require supporting research data. 



\end{document}